# Tunneling Spectroscopy of Graphene using Planar Pb Probes

Yanjing Li, Nadya Mason

*Department of Physics and Frederick Seitz Materials Research Laboratory*

*University of Illinois at Urbana-Champaign, Urbana, Illinois, 61801, United States*

**We show that evaporating lead (Pb) directly on graphene can create high-yield, high-quality tunnel probes, and we demonstrate high magnetic field/low temperature spectroscopy using these probes. Comparisons of Pb, Al and Ti/Au probes shows that after oxidation a well-formed self-limited tunnel barrier is created only between the Pb and the graphene. Tunneling spectroscopy using the Pb probes manifests energy-dependent features such as scattering resonances and localization behavior, and can thus be used to probe the microscopic electronics of graphene.**

Tunneling spectroscopy is an important method used to study the electronic properties of materials, as it has the capability of probing both the electronic density of states (DOS) and energy distributions. In particular, scanning tunneling spectroscopy (STS) has been widely used to elucidate the novel electronic properties of graphene[1], where, for example, spatial charge inhomogeneity[2, 3], Landau levels[4-7], edge states[8, 9], and phonon induced behavior[10] have been demonstrated. STS is a powerful tool, but it typically requires specialized equipment (i.e., scanning tunneling microscopes) having limited ranges of temperature, magnetic field, sample size, and surface properties. More accessible tunneling spectroscopy can be achieved via lithographically-fabricated planar tunnel junctions[11-14], where a metal probe is separated from the material under study by a thin insulating barrier. While planar tunneling cannot provide data with spatial resolution equivalent to scanning tunneling microscopy, planar tunnel probes can have similar energy resolution, and, crucially, better configurability within a standard measurement set-up (e.g., insulating substrates at ultra-low temperatures and high fields). Also, it is simple to fabricate planar tunnel probes out of various materials; for example, using a superconductor as the tunnel probe, Andreev bound states[15] and electron-electron interactions[16] can be studied. However, there has been limited use of planar tunnel junctions in graphene, likely because of the difficulty in creating the thin insulating barriers required. The common oxide insulators often do not coat graphene well; for example, aluminum oxide deposited via atomic layer deposition does not coat uniformly without extra chemical fuctionalization[17-19]. Thin $SiO_2$ on graphene typically has conducting pinholes, while metals which can be oxidized, such as Al, do not form thin uniform surface layers. Planar Cu has recently been used to create tunnel probes[20], but these require acid erosion and long time aging, and have low device yields. In contrast, we find that Pb deposited directly on graphene forms robust, high quality and high yield tunnel junctions. Here, we demonstrate that the Pb-graphene interface becomes more resistive over several days, likely due to oxidation at the interface, and that this process of oxidation is self-limiting. We compare Pb-graphene junctions to similarly-fabricated junctions of Al-graphene and Au-graphene, and show that the Pb-graphene junctions are the most reliable for reproducible tunneling experiments. Finally, we demonstrate the flexibility of the Pb tunnel junctions by utilizing them for low-temperature, magnetic and electric field-dependent spectroscopy of the graphene, observing energy-dependent features such as scattering resonances and localization behavior[21]. This shows that Pb tunnel junctions can be a simple and useful tool for attaining high energy resolution spectroscopy in graphene under a wide range of conditions.

Figure 1a shows a typical device, which consists of graphene having nearly Ohmic Ti/Au contacts at the ends, and several tunnel probes connected to the middle. The graphene was obtained using standard "scotch tape" exfoliation[22, 23] onto silicon substrates coated with 300 nm of $SiO_2$. The graphene used for this study ranged from 2-4 layers (measured via atomic force microscopy); the behavior of the probes was independent of the exact layer thickness. Standard electron-beam (e-beam) lithography and subsequent e-beam deposition were used to pattern the 4 nm Ti/700 nm Au end contacts. A second e-beam lithography step was then used to pattern the probes, which were approximately 200 nm wide and extended 100 ~ 500 nm into the graphene. Al and Ti/Au for the probes were e-beam evaporated, and the Pb was thermally evaporated. The Pb probes were covered with a 30 nm In cap to reduce oxidation at the top surface. A schematic of the finished device is shown in Figure 1b.

We studied 14 Pb probes on graphene (on 2 substrates), 7 Al probes on graphene (1 substrate), 12 Ti/Au probes on graphene (2 substrates) and 3 Pb probes on Au (1 substrate). The resistance of these probes as a function of time left under ambient conditions (i.e., oxidation time) was monitored for about 10 days starting from immediately after the device fabrication was finished. Resistance was measured in two-probe configuration by applying a bias voltage, as shown in Figure 1a; the doped substrate was used as a global backgate. Although the two-probe technique included the resistance of the Ti/Au end contact, this contact resistance was typically much lower than that of the tunnel probes (compare Figure 1c to Figure 1d), and did not have significant time-dependence. Figure 1c characterizes the typical resistance of the graphene (here, 4-layer) through the Ti/Au end contacts, as a function of backgate voltage and time. The graphene displays a clear bipolar feature and a Dirac point; the Dirac point is initially near $V_g =$ 0V and shifts toward positive voltages as time increases, likely due to oxygen doping under ambient conditions. The overall resistance of the graphene through the end contacts is relatively constant in time. This behavior can be contrasted with that of the Pb-graphene junction, as shown in Figure 1d. The Pb-graphene characteristics are initially nearly identical to those of the Ti/Au-graphene junctions, and the Dirac point again shifts toward positive voltages as oxidation time increases. However, the resistance of the Pb-graphene junctions flattens substantially over time, with the Dirac point and the bipolar features becoming less and less visible. Even more striking is the large, nearly 70-fold increase of the Pb-graphene junction resistance over time.

The increase of the Pb-graphene junction resistance is characterized in Figure 2a, which shows resistance vs. oxidation time for 5 different junctions. All the junctions demonstrate similar behavior. When first deposited, the Pb probe resistance is comparable to that of the Ti/Au contact. Then, it increases slowly (roughly doubles every half day) during the first few (1~4) days, after which it quickly (within 10 hrs) increases dramatically—by a factor of 5-10—to hundreds of k$\Omega$. After this point, the Pb resistance starts to saturate, and does not change significantly over the rest of our measurement time (an additional 6 days). The saturation resistance differs somewhat from probe to probe, even for probes on the same piece of graphene, and does not seem to correlate with parameters such as graphene thickness or probe-graphene overlap area (for the constant probe width of ~ 200nm). Similar Pb-graphene junction time evolution was observed in 11 out of 14 samples that were measured (the 3 outliers had atypical resistances for all contacts, likely due to fabrication issues). This behavior therefore seems to be highly repeatable and robust.

The large increase in Pb-graphene junction resistance with time suggests that oxidation occurs at the interface between the graphene and the Pb, which likely causes a PbO tunnel barrier to form. While it is common for oxides to form on exposed surfaces, it is less expected for oxides to form at the interfaces between materials. In the case of Pb on graphene, the interface oxidation seems to depend on the presence of graphene: in Figure 2b we show that similar Pb probes fabricated on Au pads have low resistances that do not change with time. It is possible that the weak coupling between Pb and graphene allows oxygen to diffuse into the interface from the sides[24, 25], in a way similar to the capillary action which preferentially draws etchant under graphene[26, 27]. The rate of oxidation has been found to be greatly reduced once a certain oxide thickness has been reached in the surface investigations of oxidation of bulk Pb[28], which might explain the saturation behavior. It is also possible that oxygen diffuses from the $SiO_2$ substrate through the graphene, similar to what has been seen for Pb on $YBa_2Cu_3O_7$[29]. However, this scenario seems unlikely, as it is known that graphene is highly impermeable to gases[30].

As a comparison for the Pb junctions, we also studied junctions of Au and Al on graphene. In contrast to the Pb probes, the Al probes are generally unreliable: out of seven Al probes fabricated on graphene, only two were conducting. In addition, as can be seen from Figure 2c, the resistance of the surviving probes quickly increased over time without saturating. Although Al is expected to form an $AlO_x$ tunneling barrier at the interface[20], the fast-rising resistance and low yield makes it inferior to Pb probes made in the same way. The Ti/Au probes on graphene were also less reliable than the Pb ones: 5 out of 12 Ti/Au probes on graphene were not conducting. The remaining probes had highly variable contact resistance, ranging from a few tens of kΩ to MΩ. However, as can be seen in Figure 2d, none of the Ti/Au probes showed significant time dependence of the resistance, which implies that oxidation did not occur and a tunnel barrier was not formed at the interface.

The high quality of the Pb probes is determined via transport measurements. It is particularly useful to measure below the Pb superconducting transition temperature of 7.2K, as the low voltage-bias conductance is then dominated by the characteristic Bardeen-Cooper-Schrieffer shape of the density of states. In this case, there are no single-particle states for an energy scale of $\pm\Delta$ (the superconducting gap energy, below which electrons are bound into Cooper pairs), and sharp peaks appear in the differential conductance at the gap edges. This gap feature in conductance is a typical characteristic of superconductor-normal tunneling[31], and the quality of the tunnel barrier can be determined by the quality of the gap. In particular, no conductance observed around zero bias implies that the tunnel barrier is not leaky. A lack of conduction inside the gap also implies that the tunnel barrier is fully insulating as an overly-conducting tunnel barrier would allow quasiparticle transfer inside the superconducting gap[32] (via a process known as the Andreev reflection). Figure 3 shows differential conductance vs. tunnel voltage bias for Pb-graphene junctions at 250 mK. Most Pb probes that reached a minimum resistance of several hundred k$\Omega$ showed similar behavior. The measurements were performed in 2-point configuration using standard lock-in techniques in a He-3 refrigerator, using an ac excitation of 0.02mV. In Figure 3a, the differential conductance for probes at different stages of oxidation is characterized. Probes that have not been fully oxidized (i.e., measured well before they reached saturation resistance) show no gap or incomplete gaps, and no or broad peaks at the gap edges. In contrast, the probe that was fully oxidized (black curve in Figure 3a) exhibits the full expected gap for Pb, $2\Delta=2.6$ meV, as well as well-defined peaks at the gap edges. No sub-gap conductance is evident within the resolution of our measurements, implying that the barrier is fully insulating. It is also evident in Figure 3a that a range of resistances can enable good tunneling properties, as long as a minimum resistance (presumably related to oxide thickness) is reached. Tunneling can occur for very high resistances, but the measurement may then be limited by low tunnel currents. In Figure 3b we show the magnetic field dependence of the gap, which is consistent with what is expected for superconducting tunnel junctions, where the de-pairing effect of a magnetic field lifts the gap.

With the creation of a high-quality planar tunnel junction on graphene, it is possible to perform tunneling spectroscopy at low-temperatures, and finite magnetic fields and gate voltages. Tunneling spectroscopy of graphene at ultra-low temperature and high magnetic fields has only recently been demonstrated in the most advanced STM experiments[4] and these experiments did not include a gate-voltage. Figure 4 shows 2D plots of differential conductance vs. tunnel voltage bias for a bilayer graphene sample, as a function of backgate voltage (a,b) and magnetic field (c). For this data, we focus on energy scales much larger than the superconducting gap energy (i.e., where the Pb probe is normal). In the data, patterns of peaks and oscillations are evident as white lines (of high differential conductance) that evolve with tunnel bias, magnetic field, and/or backgate voltage. Although a detailed analysis of all these features is beyond the scope of this paper, here we describe the likely origins of the main features, to demonstrate the quality and potential of the tunnel probes. In the backgate-dependent spectra of Figure 4a, the arrow points to fine oscillations that evolve diagonally away from the zero-bias gap towards higher biases. These oscillations disappear with applied magnetic field (compare Figure 4b), which implies that they are related to superconductivity in the probe. In particular, these features are similar to McMillan-Rowell oscillations, which are due to interference between quasiparticles reflected from superconductor-normal interfaces[33], and are evident via tunneling measurements. Such oscillations can be used to extract parameters such as the Fermi velocity $v_F$, as they are predicted to have bias spacing of $\Delta V_b = hv_F/(4ed_N)$, where $d_N$ is the length scale of the normal metal[34]; for our experiments, although $d_N$ varies in the 2D graphene, the predicted $\Delta V_b \sim 0.5$mV is comparable to the typical peak spacing.

An additional set of broader resonances are marked in Figure 4a and 4b by white lines; these peaks do not disappear with applied magnetic field, implying that they are not related to superconductivity. In fact, these resonances are very similar to features recently seen in STM studies[21], which were attributed to weakly localized states in graphene created by impurity scattering or substrate-induced disorder potentials. Near zero-bias, Coulomb charging is evident by diamond-like structures (circled in Figure 4b); in this case, the localization may occur due to interference between random scattering events within a phase coherence length[21, 35]. In Figure 4c, diamond-like features are also evident in the magnetic-field dependence of the differential conductance. These interference features are reminiscent of Fabry-Perot oscillations, and may be due to interference between magnetic field-induced edge states, which have been separated into conducting and insulating strips by local disorder[36]. Future studies, at higher magnetic fields and sample mobilities[7], will determine the evolution of Landau levels in the samples.

In conclusion, we have found a simple, efficient and reliable way of making planar tunneling probes by directly depositing Pb on graphene. The oxidation of the Pb-graphene interface seems uniquely self-limiting. We have also performed tunneling spectroscopy using these probes and observed features such as scattering resonances and localization. Our findings provide new possibilities for investigating the microscopic electronic properties of graphene via tunneling spectroscopy.

The manuscript was written through contributions of all authors. All authors have given approval to the final version of the manuscript. We thank M.J. Gilbert for useful discussions. This work was supported by the NSF DMR-0906521.

Corresponding Author: N. Mason (nadya@illinois.edu)

REFERENCES

1. Andrei, E. Y.; Li, G.; Du, X. *Rep. Prog. Phys.* **2012,** 75, 056501.
2. Martin, J.; Akerman, N.; Ulbricht, G.; Lohmann, T.; Smet, J. H.; von Klitzing, K.; Yacoby, A. *Nat. Phys.* **2008,** 4, 144-148.
3. Zhang, Y.; Brar, V. W.; Girit, C.; Zettl, A.; Crommie, M. F. *Nat. Phys.* **2009,** 5, 722-726.
4. Song, Y. J.; Otte, A. F.; Kuk, Y.; Hu, Y.; Torrance, D. B.; First, P. N.; de Heer, W. A.; Min, H.; Adam, S.; Stiles, M. D.; MacDonald, A. H.; Stroscio, J. A. *Nature* **2010,** 467, 185-189.
5. Miller, D. L.; Kubista, K. D.; Rutter, G. M.; Ruan, M.; de Heer, W. A.; First, P. N.; Stroscio, J. A. *Science* **2009,** 324, 924-927.
6. Li, G.; Luican, A.; Andrei, E. Y. *Phys. Rev. Lett.* **2009,** 102, 176804.
7. Luican, A.; Li, G.; Andrei, E. Y. *Phys. Rev. B* **2011,** 83, 041405.
8. Tao, C.; Jiao, L.; Yazyev, O. V.; Chen, Y.-C.; Feng, J.; Zhang, X.; Capaz, R. B.; Tour, J. M.; Zettl, A.; Louie, S. G.; Dai, H.; Crommie, M. F. *Nat. Phys.* **2011,** 7, 616-620.
9. Kobayashi, Y.; Fukui, K.-i.; Enoki, T.; Kusakabe, K. *Phys. Rev. B* **2006,** 73, 125415.
10. Zhang, Y.; Brar, V. W.; Wang, F.; Girit, C.; Yayon, Y.; Panlasigui, M.; Zettl, A.; Crommie, M. F. *Nat. Phys.* **2008,** 4, 627-630.
11. Amet, F.; Williams, J. R.; Garcia, A. G. F.; Yankowitz, M.; Watanabe, K.; Taniguchi, T.; Goldhaber-Gordon, D. *Phys. Rev. B* **2012,** 85, 073405.
12. Staley, N.; Wang, H.; Puls, C.; Forster, J.; Jackson, T. N.; McCarthy, K.; Clouser, B.; Liu, Y. *Appl. Phys. Lett.* **2007,** 90, 143518.
13. Zeng, C.; Wang, M.; Zhou, Y.; Lang, M.; Lian, B.; Song, E.; Xu, G.; Tang, J.; Torres, C.; Wang, K. L. *Appl. Phys. Lett.* **2010,** 97, 032104.
14. Vora, H.; Kumaravadivel, P.; Nielsen, B.; Du, X. *Appl. Phys. Lett.* 100, 153507-5.
15. Dirks, T.; Hughes, T. L.; Lal, S.; Uchoa, B.; Chen, Y.-F.; Chialvo, C.; Goldbart, P. M.; Mason, N. *Nat. Phys.* **2011,** 7, 386-390.
16. Chen, Y.-F.; Dirks, T.; Al-Zoubi, G.; Birge, N. O.; Mason, N. *Phys. Rev. Lett.* **2009,** 102, 036804.
17. Lee, B.; Park, S.-Y.; Kim, H.-C.; Cho, K.; Vogel, E. M.; Kim, M. J.; Wallace, R. M.; Kim, J. *Appl. Phys. Lett.* **2008,** 92, 203102.
18. Wang, X.; Tabakman, S. M.; Dai, H. *J. Am Chem Soc.* **2008,** 130, 8152-8153.
19. Wang, L.; Travis, J. J.; Cavanagh, A. S.; Liu, X.; Koenig, S. P.; Huang, P. Y.; George, S. M.; Bunch, J. S. *Nano lett.* **2012,** 12, 3706-3710.
20. Malec, C. E.; Davidović, D. *J Appl Phys.* **2011,** 109, 064507.
21. Jung, S.; Rutter, G. M.; Klimov, N. N.; Newell, D. B.; Calizo, I.; Hight-Walker, A. R.; Zhitenev, N. B.; Stroscio, J. A. *Nat. Phys.* **2011,** 7, 245-251.
22. Novoselov, K. S.; Geim, A. K.; Morozov, S. V.; Jiang, D.; Katsnelson, M. I.; Grigorieva, I. V.; Dubonos, S. V.; Firsov, A. A. *Nature* **2005,** 438, 197-200.
23. Miao, F.; Wijeratne, S.; Zhang, Y.; Coskun, U. C.; Bao, W.; Lau, C. N. *Science* **2007,** 317, 1530-1533.
24. Malec, C. E.; Davidović, D. *Phys. Rev. B* **2011,** 84, 121408.
25. Li, S.-L.; Miyazaki, H.; Kumatani, A.; Kanda, A.; Tsukagoshi, K. *Nano lett.* **2010,** 10, 2357-2362.
26. Du, X.; Skachko, I.; Barker, A.; Andrei, E. Y. *Nat. Nanotechnol.* **2008,** 3, 491-495.
27. Bolotin, K. I.; Sikes, K. J.; Jiang, Z.; Klima, M.; Fudenberg, G.; Hone, J.; Kim, P.; Stormer, H. L. *Solid State Commun.* **2008,** 146, 351-355.
28. Peeters, F.; Slavin, A. J. *Surf. Sci.* **1989,** 214, 85-96.

29. Covington, M.; Scheuerer, R.; Bloom, K.; Greene, L. H. *Appl. Phys. Lett.* **1996,** 68, 1717-1719.
30. Bunch, J. S.; Verbridge, S. S.; Alden, J. S.; van der Zande, A. M.; Parpia, J. M.; Craighead, H. G.; McEuen, P. L. *Nano lett.* **2008,** 8, 2458-2462.
31. Tinkham, M., *Introduction to Superconductivity*. 2nd ed.; Dover: New York, 2004; pp 75-76.
32. Blonder, G. E.; Tinkham, M.; Klapwijk, T. M. *Phys. Rev. B* **1982,** 25, 4515-4532.
33. Rowell, J. M.; McMillan, W. L. *Phys. Rev. Lett.* **1966,** 16, (11), 453-456.
34. Nesher, O.; Koren, G. *Phys. Rev. B* **1999,** 60, 9287-9290.
35. Rutter, G. M.; Crain, J. N.; Guisinger, N. P.; Li, T.; First, P. N.; Stroscio, J. A. *Science* **2007,** 317, 219-222.
36. Velasco, J., Jr.; Liu, G.; Jing, L.; Kratz, P.; Zhang, H.; Bao, W.; Bockrath, M.; Lau, C. N. *Phys. Rev. B* 81, 121407.

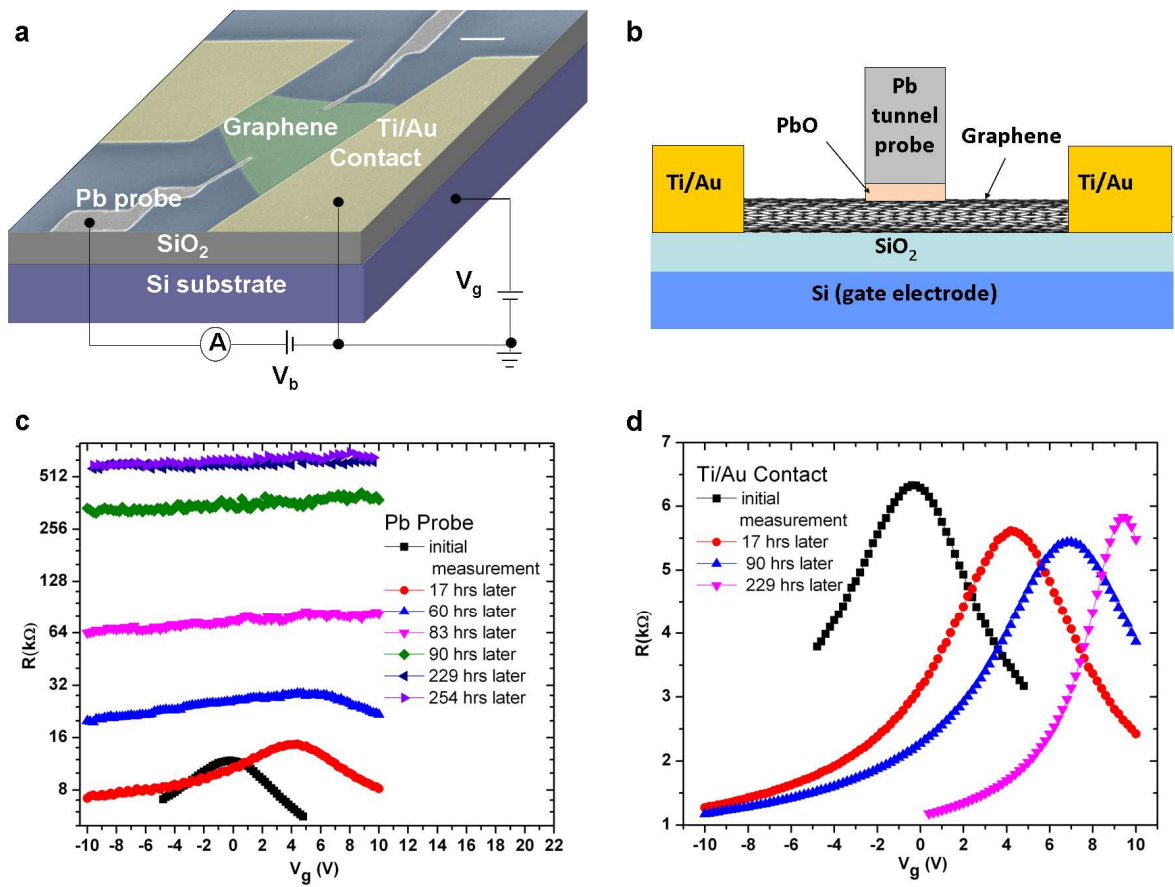

**Figure 1.** Illustration of device and electronic characterization at room temperature. **(a)** Scanning electron micrograph (SEM) of a typical device with an overlaid measurement circuit (false-colored for improved contrast). The scale bar is 2 μm. **(b)** Side-view schematic diagram of the device, showing where PbO forms under ambient conditions. **(c)** Typical resistance vs. backgate voltage for Ti/Au-Ti/Au end contacts on graphene, as a function of time spent under ambient conditions. Initial measurement is performed ~1hr after the device fabrication is finished. **(d)** Typical resistance vs. backgate voltage for a Pb probe on graphene, as a function of time spent under ambient conditions.

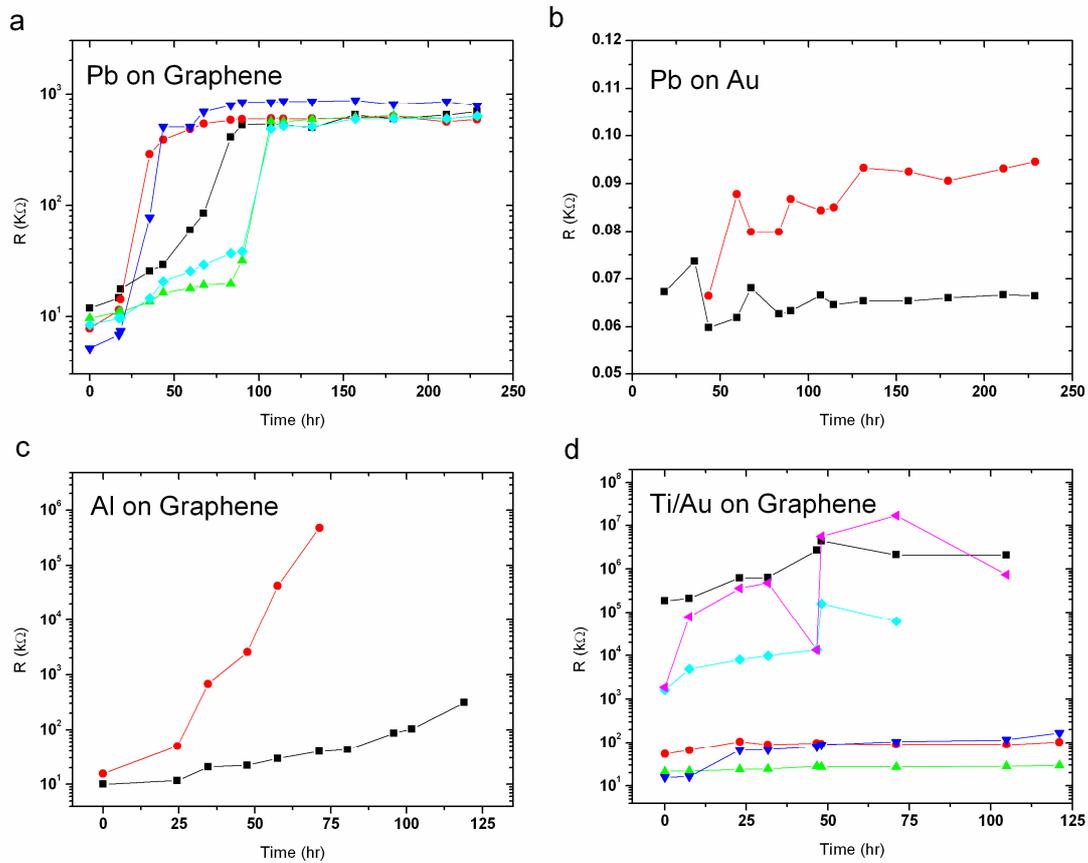

**Figure 2.** Distinct behaviors of probes made from Pb, Al and Ti/Au on graphene, and of Pb on Au. Time dependence (i.e., wait time under ambient conditions) of the probe-on-graphene resistance for different probe materials: **(a)** Pb , **(c)** Al, and **(d)** Ti/Au. **(b)** Resistance of Pb probe on Au pad as a function of time. Different colors represent data for different probes (for each material, probes were fabricated on several different graphene flakes, on multiple substrates).

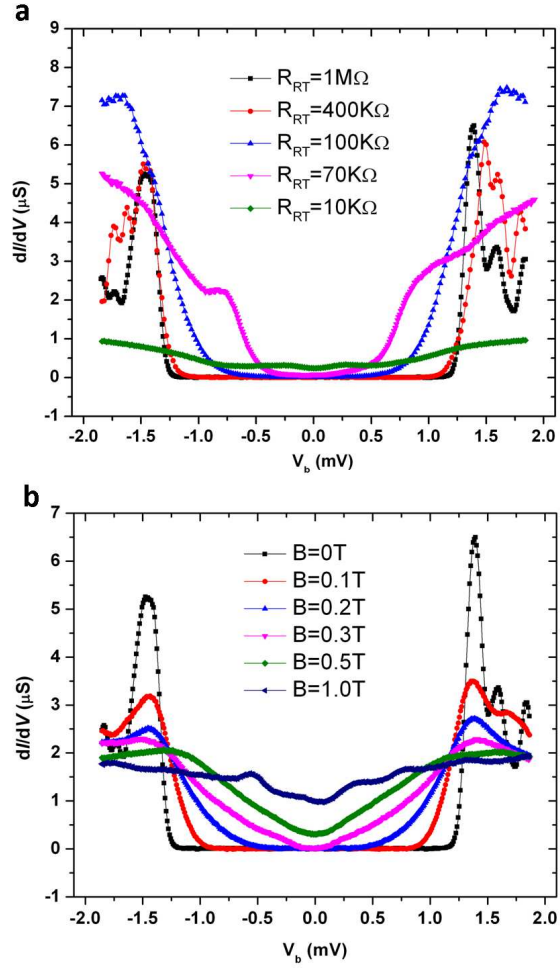

**Figure 3.** Characterization of the Pb tunneling probe at 250 mK. **(a)** Differential conductance vs. tunnel bias $V_b$ for Pb probes having different resistances at room temperature (oxidized for different times). Conductance for different probes is normalized according to their room temperature resistances. The more resistive probes show a cleaner gap feature, which originates from superconducting DOS. Oscillations outside the gap edge are shown as fine oscillations in Figure 4a. **(b)** Magnetic field dependence of differential conductance vs. bias voltage. As magnetic field increases, the gap feature gradually disappears, confirming that the gap feature arises from superconductivity.

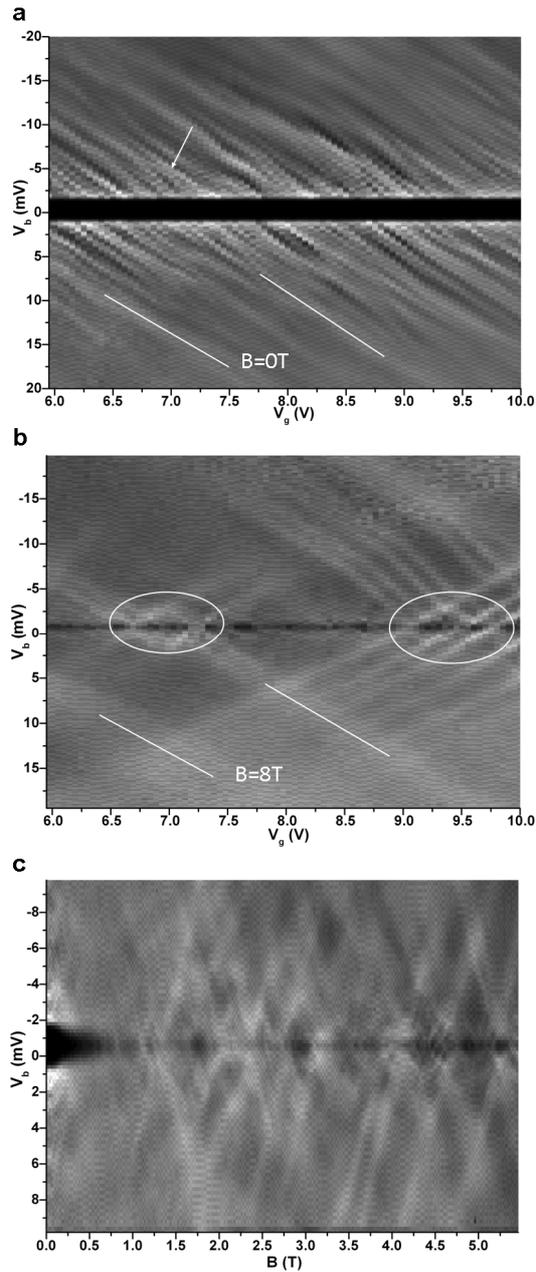

**Figure 4.** Gate and magnetic-field dependence of tunneling spectra. **(a)** Two-dimensional (2D) map of differential conductance versus dc bias voltage $V_b$ and backgate voltage $V_g$, taken on a bilayer sample. The dark horizontal band centered at zero bias comes from the superconducting gap. Arrows point to fine oscillations that disappear with applied magnetic field. White lines are guides to the eye showing broader resonances which persist with magnetic field. **(b)** Similar 2D map as (a) except with a magnetic field of 8T perpendicular to the plane of substrate. White lines show broad resonances similar to those in (a). Circles show regions of Coulomb blockade diamonds. **(c)** 2D map of differential conductance versus $V_b$ and magnetic field $B$ taken at zero backgate voltage. Note smaller bias voltage scale in (c).

14